\documentclass[a4paper,12pt]{article}
\usepackage{geometry}
\usepackage[cp1251]{inputenc}
\usepackage{epsfig}
\usepackage{graphicx}

\usepackage{amssymb}

\begin{document}

Astronomy Letters, 2013, Vol. 39, No. 5, pp. 0-00 \hspace{20pt} 
Printed 16 April 2013

\vspace{-8pt}

\section*{
\begin{center}
Magnetically Active Stars in Taurus--Auriga: Evolutionary Status
\end{center}
}

\begin{center}
\Large K. N. Grankin
\end{center}

\begin{center}
\it{Crimean Astrophysical Observatory, Nauchny, Crimea, 98409 Ukraine}
\end{center}

\begin{center}
konstantin.grankin@rambler.ru
\end{center}

\subsection*{\center Abstract}

\begin{quote}
We have analyzed a sample of 74 magnetically active stars toward the 
Taurus--Auriga star-forming region. Based on accurate data on their 
basic physical parameters obtained from original photometric 
observations and published data on their proper motions, X-ray 
luminosities, and equivalent widths of the $\rm H\alpha$ and Li lines,
we have refined the evolutionary status of these objects. We show that
50 objects are young stars with ages of 1-40 Myr and belong to the
Taurus--Auriga star-forming region. Other 20 objects have a 
controversial evolutionary status and can belong to both Taurus--Auriga 
star-forming region and the Gould Belt. The remaining four objects with 
ages of 70-100 Myr belong to the zero-age main sequence. We have 
analyzed the relationship between the rotation period, mass, and age
for 50 magnetically active stars. The change in the angular momentum of 
the sample stars within the first 40 Myr of their evolution has been 
investigated. An active star-protoplanetary disk interaction is shown to
occur on a time scale from 0.7 to 10 Myr.
\end{quote}

\vspace{15pt}

\textbf{Key words:} {\it stars -- variable, properties, rotation, 
pre-main-sequence stars.}

\subsection*{INTRODUCTION}
\indent

The Taurus--Auriga star-forming region (SFR)
includes a rich population of young pre-main-sequence
(PMS) stars. According to Kenyon et al. (2008), the
most complete sample of PMS stars in the Taurus--Auriga
SFR contains 383 objects. This sample is
represented by infrared objects or protostars, which
are optically invisible because of strong absorption
in the surrounding gas-dust clouds, T Tauri stars
(TTS), and brown dwarfs. In turn, the subgroup of
T Tauri stars includes both classical TTS (CTTS)
with strong emission lines and significant infrared
(IR) and ultraviolet (UV) excesses and weak-lined
TTS (WTTS) with insignificant IR and UV excesses
or without them. To all appearances, CTTS are at
the stage of an active interaction with their accretion
disks, while the disks around WTTS are either absent
or not accretion ones.

The listed objects form a rough age sequence with
protostars as the youngest objects and WTTS as the
oldest ones. The youngest protostars are concentrated
in the densest parts of the molecular clouds;
CTTS are located in both the clouds themselves and
in the narrow dark bands that connect the largest
clouds. WTTS show a lesser concentration to the
clouds and can be slightly older than CTTS. Most of
the PMS stars in the Taurus--Auriga SFR have ages
of 1-3 Myr and only some of the SFR members have
ages older than 3 Myr. If we take into account the fact
that the molecular clouds in the SFR exist for about
10 Myr or more and assume that the star formation
rate remained constant during the entire lifetime of
the clouds, then we have to admit that the known
PMS stars must represent only a small part of this
association.

The absence of an observable population of older
PMS stars with ages of about 3 Myr or older is characteristic
not only for the Taurus--Auriga SFR but
also for almost all of the neighboring SFRs (see, e.g.,
Carpenter 2000; Palla and Stahler 2000). Either star
formation takes place during the short lifetime of the
molecular clouds or the molecular clouds themselves
exist only for several Myr. This problem has been
discussed in the literature for almost three decades
and is known as the problem of the absence of older
PMS stars that have recently passed the TTS stage
(post T Tauri stars or PTTS) (Herbig 1978; Feigelson
1996).

All of the stars later than the spectral type F and
younger than 100 Myr manifest themselves as soft
X-ray sources and they can be easily detected with
the help of sensitive X-ray observatories. Several
studies were undertaken to find new candidates for
PMS stars among the multitude of X-ray sources toward
the Taurus--Auriga SFR (see, e.g., Wichmann
et al. 1996; Neuh\"auser et al. 1997). These studies
revealed a significant population of candidates for
PMS stars that are widely dispersed over the Taurus--Auriga
SFR and beyond. Since young stars exhibit small changes in 
their X-ray properties within the first 100 Myr of their 
evolution, the X-ray data are insufficient to distinguish PMS 
stars with ages of 3-50 Myr
from zero-age main-sequence (ZAMS) stars with
ages of about 100 Myr. Thus, the sample of X-ray active
stars can be heavily “contaminated” with field
objects and additional spectroscopic observations are
required to identify new PMS stars.

Using data from the ROSAT all-sky survey and
spectroscopic data, Wichmann et al. (1996) investigated
a region with a size of about 280 sq. degrees
in the Taurus--Auriga SFR. As a result of this
work, 76 candidates for PMS stars were identified
(below referred to as Wichmann’s list). Wichmann
et al. (1996) used the following criteria to classify the
stars as PMS: the presence of the Li I $\lambda$6707 absorption
line with an equivalent width $\geq$ 100 m\AA\ and
the spectral type F or later. Most of these stars (72)
were classified as WTTS based on the detection of
weak emission or absorption in the $\rm H\alpha$ line 
(EW($\rm H\alpha$) $\leq$ 10\AA). The remaining six PMS stars 
with strong $\rm H\alpha$ emission were classified as CTTS.

Martin and Magazz\`u (1999) disagreed with the
classification criteria adopted by Wichmann et al.
(1996). They pointed out that the detection of lithium
in a late-type star is a necessary but not sufficient
condition for it to be regarded as WTTS, because
PTTS and low-mass members of young open clusters
with ages of 30-200 Myr also show strong lithium
lines (see Martin 1997). Martin and Magazz\`u (1999)
used higher-resolution spectra for 35 stars from
Wichmann’s list to measure the equivalent widths of
the lithium line and to revise the evolutionary status
of these stars. They showed that their sample is a
mixture of young (11 TTS and 8 PTTS) and older
ZAMS stars. Wichmann et al. (2000) undertook
another detailed study of 58 PMS candidates from
Wichmann’s list based on high-resolution echelle
spectroscopy and proper motions. They found that
approximately 60\% of the stars from this sample
could be regarded as PMS, while the remaining
stars probably belong to the ZAMS. Wichmann
et al. (2000) concluded that the PMS stars were
probably associated with the Taurus--Auriga SFR,
while the ZAMS stars could represent the population
of older stars in the Gould Belt. Nevertheless, none of
the objects from Wichmann et al. (2000) was included
in the list of recognized members of the Taurus--
Auriga SFR published by Kenyon et al. (2008).

A long-term photometric monitoring of a representative
sample of PMS stars had been performed as
part of the ROTOR program at the Maidanak Astronomical
Observatory in Uzbekistan for almost twenty
years (1984-2006). As a result of the implementation
of this program, more than 100 000 $UBVR$ measurements
were obtained for 370 objects in various
SFRs. The final results of our photometric observations
for 72 CTTS and 48 WTTS are presented
in Grankin et al. (2007) and Grankin et al. (2008),
respectively. There were several tens of PMS stars
from the Taurus--Auriga SFR among the observed
objects: 34 CTTS and 40 WTTS. Since these stars
are recognized members of the Taurus--Auriga SFR
(see Kenyon et al. 2008), below we will call them
\textquotedblleft well-known PMS stars\textquotedblright.

In addition, 60 candidates for PMS stars from
Wichmann et al. (1996) were observed at the Maidanak
Astronomical Observatory. More than 5000
$BVR$ measurements were obtained for these stars
during 1994-2006. Since the objects from Wichmann’s
list were not included in the list of recognized
PMS stars from the Taurus--Auriga SFR published
by Kenyon et al. (2008), below we will call these
objects \textquotedblleft candidates for PMS 
stars\textquotedblright.

Previously (Grankin 2013a), we analyzed homogeneous
long-term photometric observations of
28 well-known PMS stars published in Grankin
et al. (2008) and 60 candidates for PMS stars from
Wichmann’s list toward the Taurus--Auriga SFR.
We were able to determine reliable luminosities,
radii, masses, and ages only for 74 of the 88 sample
stars. In addition, 62 stars from this sample were
shown to exhibit periodic light variations due to the
phenomenon of spotted rotational modulation.

This paper is a logical continuation of our studies
of a representative sample of stars begun in Grankin
(2013a). Our main goals are: (1) to refine
the evolutionary status of the candidates for PMS
stars and to compare it with the evolutionary status of
the well-known PMS stars from the Taurus--Auriga
SFR; (2) to analyze the possible relationship between
the rotation period, mass, and age for all sample
stars; and (3) to investigate the change in the angular
momentum of the sample stars within the first 40 Myr
of their evolution.

\subsection*{PUBLISHED DATA}
\indent

To successfully solve the formulated problems,
we collected published data on the proper motions,
$\rm H\alpha$ and Li I (6707 \AA) line equivalent widths, and
X-ray luminosities for all sample stars. The proper
motions were taken from Ducourant et al. (2005)
and are given in columns 2 and 3 of Table 1. The
data on the Li I (6707 \AA) and $\rm H\alpha$ line equivalent
widths were compiled from the following papers:
Herbig et al. (1986), Hartmann et al. (1987),
Walter et al. (1988), Strom et al. (1989), Gomez
et al. (1992), Magazz\`u et al. (1992), Martin et al.
(1994), Martin and Magazz\`u (1999), Wichmann
et al. (2000), Basri et al. (1991), and Nguyen
et al. (2009). The median Li I (6707 \AA) and $\rm H\alpha$
line equivalent widths are given in columns 4 and 5
of Table 1. The X-ray luminosities were taken from
Stelzer and Neuh\"auser (2001) and G\"udel et al.
(2007) and are given in the last column. Table 2
presents analogous data for the objects from our
previous paper (Grankin 2013a) that, for one reason
or another, have no reliable data on their luminosities,
radii, masses, or ages.

\begin{table}[t]
\caption{Published data for the sample stars}
\centering 
\vspace{5mm}
\begin{footnotesize}
\tabcolsep=0.25em
\begin{tabular}{c|c|c|r|r|c||r|c|c|r|r|c}
\hline\hline
\rule{0pt}{2pt}&&&&&&&&&&\\
W96 & $\mu_\alpha\cos \delta$ & $\mu_\delta$ & W(Li) & $\rm W(H\alpha)$ & Lx, $10^{30}$ & 
W96/name & $\mu_\alpha\cos \delta$ & $\mu_\delta$ & W(Li) & $\rm W(H\alpha)$ & Lx, $10^{30}$ \\
       & $\mbox{mas/yr}$ & $\mbox{mas/yr}$ & \makebox[1.6em][l]{\AA} & 
\makebox[2.2em][l] {\AA} & $\mbox{erg s}^{-1}$ & & $\mbox{mas/yr}$ & 
$\mbox{mas/yr}$ & \makebox[1.6em][l] {\AA} & \makebox[2.2em][l] { \AA} & 
$\mbox{erg s}^{-1}$ \\[3pt]
\hline 
\rule{0pt}{5mm}
 01 &  2 & -11 &  0.26 &  0.38 & 2.94 &       60 &     &     & 0.10 &  1.50 & 0.37 \\
 02 &  3 & -14 &  0.45 & -0.25 & 1.06 &       62 &   0 & -18 & 0.40 & -0.05 & 0.72 \\
 03 & 16 & -20 &  0.22 &  0.70 & 2.31 &       63 &   9 & -20 & 0.58 & -0.50 & 1.11 \\
 04 &  4 & -17 &  0.35 &  0.64 & 4.58 &       64 &  -2 & -17 & 0.31 & -0.29 & 1.11 \\
 05 &  6 & -15 &  0.25 &  2.77 & 3.46 &       66 &  -1 & -18 &      &  2.44 & 4.37 \\
 06 &  7 & -15 &  0.21 &  3.28 & 3.14 &       67 &   0 & -17 & 0.25 &  0.60 & 0.97 \\
 07 &  7 & -21 &  0.59 & -4.60 & 3.54 &       68 &  11 & -27 &      &  0.10 & 0.48 \\
 08 & 18 & -53 &  0.26 & -0.20 & 2.31 &       70 &  12 & -18 & 0.30 & -7.00 & 1.45 \\
 10 & 20 & -32 &  0.41 &  1.00 & 3.21 &       71 &   0 & -18 & 0.20 &  2.71 & 0.74 \\
 11 &  3 & -11 &  0.30 & -3.70 & 1.35 &       73 &  11 & -17 & 0.45 &  0.30 & 0.94 \\
 12 & 30 & -43 &  0.26 & -1.80 & 0.74 &       74 &   2 & -18 & 0.43 & -0.10 & 3.49 \\
 13 & -1 & -18 &  0.41 & -0.35 & 1.33 &       75 &   8 & -25 & 0.44 &  0.03 &      \\
 14 &  8 & -16 &  0.17 &  1.10 & 0.82 &       76 &  12 & -18 & 0.40 & -0.06 &  3.3 \\
 15 & 16 & -40 &  0.16 &  2.60 & 0.40 &   Anon 1 &     &     & 0.48 & -2.50 & 1.45 \\
 18 &  1 & -14 &  0.27 &  1.88 & 2.46 &   TAP 31 &   8 & -27 &      &  1.01 & 9.12 \\
 23 & -7 & -22 &  0.37 & -0.57 & 0.47 &   LkCa 1 &  16 & -29 & 0.56 & -3.40 & 0.11 \\
 27 &  8 & -16 &  0.36 &  0.75 & 2.93 &   LkCa 4 &   8 & -24 & 0.61 & -4.05 & 0.91 \\
 29 &  2 & -25 &  0.34 &  1.35 & 4.99 &   LkCa 5 &   7 & -33 & 0.55 & -3.90 & 0.34 \\
 30 &  1 & -17 &  0.16 &  2.73 &      &   LkCa 7 &   5 & -31 & 0.59 & -3.70 & 0.83 \\
 31 & 29 & -28 &  0.14 &  1.20 & 3.53 &  LkCa 14 &   4 & -21 & 0.60 & -0.90 & 1.07 \\
 32 &  3 & -14 &       &  1.39 & 1.87 &  LkCa 16 &  10 & -17 & 0.44 & -4.00 & 1.12 \\
 36 & -5 & -12 &  0.24 &  0.66 & 0.20 &  LkCa 19 &   3 & -19 & 0.47 & -0.50 & 5.50 \\
 37 &  8 & -16 &  0.54 & -1.80 & 0.21 &  LkCa 21 &  12 & -30 & 0.75 & -5.50 & 1.82 \\
 39 &-13 &  -6 &$<$0.05&  2.30 & 1.02 &    TAP 4 &  25 & -41 & 0.35 & -0.10 &      \\
 40 & 13 & -23 &  0.49 & -2.50 & 3.88 &    TAP 9 &  20 & -48 & 0.34 & -1.60 & 0.78 \\
 41 &  0 & -11 &  0.36 &  0.51 & 0.62 &   TAP 26 &   9 & -15 & 0.57 & -1.05 & 0.89 \\
 44 &  7 &  -3 &  0.35 &  0.24 & 0.44 &   TAP 35 &   2 & -15 & 0.28 &  1.35 & 2.57 \\
 45 & -8 & -16 &  0.25 & -1.30 & 0.70 &   TAP 40 &  10 & -28 & 0.15 & -0.29 & 0.38 \\
 46 &  7 & -11 &  0.38 & -0.50 & 0.29 &   TAP 41 &  11 & -18 & 0.66 & -0.53 & 0.95 \\
 47 & 14 & -19 &  0.42 & -0.07 & 1.24 &   TAP 45 &  16 & -13 & 0.60 & -0.70 & 0.35 \\
 48 & 17 & -46 &  0.50 & -0.39 & 2.72 &   TAP 50 & -51 & -12 &      & -0.60 &      \\
 53 &    &     &$<$0.07&  0.00 & 0.95 &   TAP 57 &   2 & -26 & 0.58 & -1.05 & 0.78 \\
 54 &  7 & -23 &  0.19 & -2.10 & 2.49 & V819 Tau &   9 & -32 & 0.62 & -2.50 & 0.79 \\
 56 & 41 & -15 &$<$0.05& -0.90 & 0.71 & V826 Tau &  13 & -22 &      & -3.30 & 1.35 \\
 57 & -4 &  -9 &  0.20 &  1.85 & 0.91 & V827 Tau &   8 & -15 & 0.57 & -2.95 & 1.95 \\
 58 &  1 & -18 &  0.28 &  1.40 & 2.14 & V830 Tau &  -8 & -28 & 0.68 & -2.00 & 1.91 \\
 59 & 16 & -23 &  0.47 & -5.00 & 0.90 &   VY Tau &  12 & -20 & 0.52 & -4.90 & 0.98 \\
\hline
\end{tabular}
\end{footnotesize}
\end{table}
\clearpage

\begin{table}[t]

\vspace{6mm}
\caption{Published data for the sample stars without reliable 
information about their luminosities, radii, masses, and ages}
\centering 
\vspace{5mm}
\begin{footnotesize}
\tabcolsep=0.25em
\begin{tabular}{c|c|c|r|r|c||r|c|c|r|r|c}
\hline\hline
\rule{0pt}{2pt}&&&&&&&&&&\\
W96 & $\mu_\alpha\cos \delta$ & $\mu_\delta$ & W(Li) & $\rm W(H\alpha)$ 
& Lx, $10^{30}$ & 
W96/name & $\mu_\alpha\cos \delta$ & $\mu_\delta$ & W(Li) & 
$\rm W(H\alpha)$ & Lx, $10^{30}$\\
& $\mbox{mas/yr}$ & $\mbox{mas/yr}$ & \makebox[1.6em][l]{\AA} & 
\makebox[2.2em][l] 
{\AA} & $\mbox{erg s}^{-1}$ & & $\mbox{mas/yr}$ & $\mbox{mas/yr}$ & 
\makebox[1.6em][l] 
{\AA} & \makebox[2.2em][l] { \AA} & $\mbox{erg s}^{-1}$ \\[3pt]
\hline 
\rule{0pt}{5mm}
 09 & 16 & -17 &$<$0.07 & -0.10 & 2.64 &       55 & 12 & -18 &   0.36 &  1.40 & 3.66 \\
 19 &  1 &  -9 &   0.17 &  2.10 & 0.53 &       61 &    &     &   0.19 &  2.80 & 0.37 \\
 28 & 11 & -27 &   0.40 &  0.40 &      &       65 &    &     &        &  3.80 & 0.31 \\
 38 &  3 &  -7 &        &  2.10 & 0.41 &       72 & -6 & -12 &$<$0.11 &  0.90 & 1.23 \\
 49 &  4 &  -5 &   0.14 &  2.55 & 0.53 &   LkCa 3 &  6 & -21 &   0.56 & -2.70 & 0.68 \\
 50 & -2 & -21 &   0.23 &  2.88 & 0.70 & V410 Tau &  6 & -31 &   0.54 & -2.25 & 2.40 \\
 51 &  2 & -15 &        &  1.80 & 0.22 & V836 Tau & 11 & -10 &   0.57 & -7.70 & 0.74 \\
 52 & -2 &   7 &        &  1.00 & 0.52 &   TAP 49 &  0 &  -7 &   0.27 &  0.90 & 0.05 \\
\hline
\end{tabular}  
\end{footnotesize}
\end{table}

\subsection*{PROPER MOTION}
\indent

Figure 1 presents 70 objects from our sample
with available proper motions from Ducourant
et al. (2005). The well-known PMS stars and PMS
candidates are designated by the gray and black
colors, respectively. The objects with reliable physical
parameters given in Table~1 are marked by the circles;
the objects without reliable parameters presented in
Table 2 are designated by the squares. To analyze the
proper motions of the objects from our sample, we
invoked additional data on the proper motions of 123
recognized members of the Taurus--Auriga SFR from
the same paper by Ducourant et al. (2005). These are
designated by the crosses on the plot. Using these
data, we calculated the mean and standard deviation
for the proper motions of the recognized members of
the Taurus--Auriga SFR:
\\

$\mu_\alpha cos\delta = \quad \; 5.83 \pm 5.91$  mas $\rm yr^{-1}$, 

$\qquad \mu_\delta =  -19.52 \pm 7.58$  mas $\rm yr^{-1}$.\\

The large ellipse in Fig. 1 bounds the 3$\sigma$ region
around this mean. It can be seen from the figure
that most of the stars from our sample are within
this ellipse, i.e., their proper motions correspond to
those of the recognized SFR members. At the same
time, there are several objects, both among the stars
of our sample and among the recognized SFR members,
whose proper motions differ significantly from
the mean proper motion of the SFR. Seven objects
should be pointed out among the recognized SFR
members from Kenyon’s list: HBC~355, HBC~358,
HBC~360, HBC~361, HBC~362, HBC~375, and
HBC~418. Our list contains 10 such objects: W39,
W31, W56, W12, W08, W48, W15, TAP~4, TAP~9,
and TAP~50; 5 of these 10 stars (W08, W48, W15,
TAP~4, and TAP~9) have proper motions close to
those of the Pleiades members (enclosed in the small
gray circle). Similar conclusions about the kinematic
properties of these five stars from our sample were
reached by Frink et al. (1997). In particular, they
point out that TAP~4, TAP~9, W08, W15, and W48
are kinematic members of the Pleiades.

A detailed kinematic study of the Taurus--Auriga
SFR was undertaken by Bertout and Genova (2006).
They also determined the kinematic parallaxes for 15
PMS candidates from Wichmann’s list and found
that 8 of these 15 candidates are SFR members
(W05, W06, W27, W32, W37, W62, W71, W74)
and 7 others are not (W01, W07, W14, W29, W54,
W58, W67). However, all these 15 stars have proper
motions very close to the mean proper motion of the
SFR (the deviations from the mean are less than 1$\sigma$).
Therefore, we believe that all 15 stars are possible
members of the Taurus--Auriga SFR. It should be
noted that the proper motions of 13 objects from our
sample without reliable data on their luminosities,
radii, masses, and ages agree well with the mean
proper motion of the SFR within 3$\sigma$ (see Table 2).
Only one star, W52, constitutes an exception.

\begin{figure}[ht]
\epsfxsize=11cm
\vspace{0.6cm}
\hspace{2cm}\epsffile{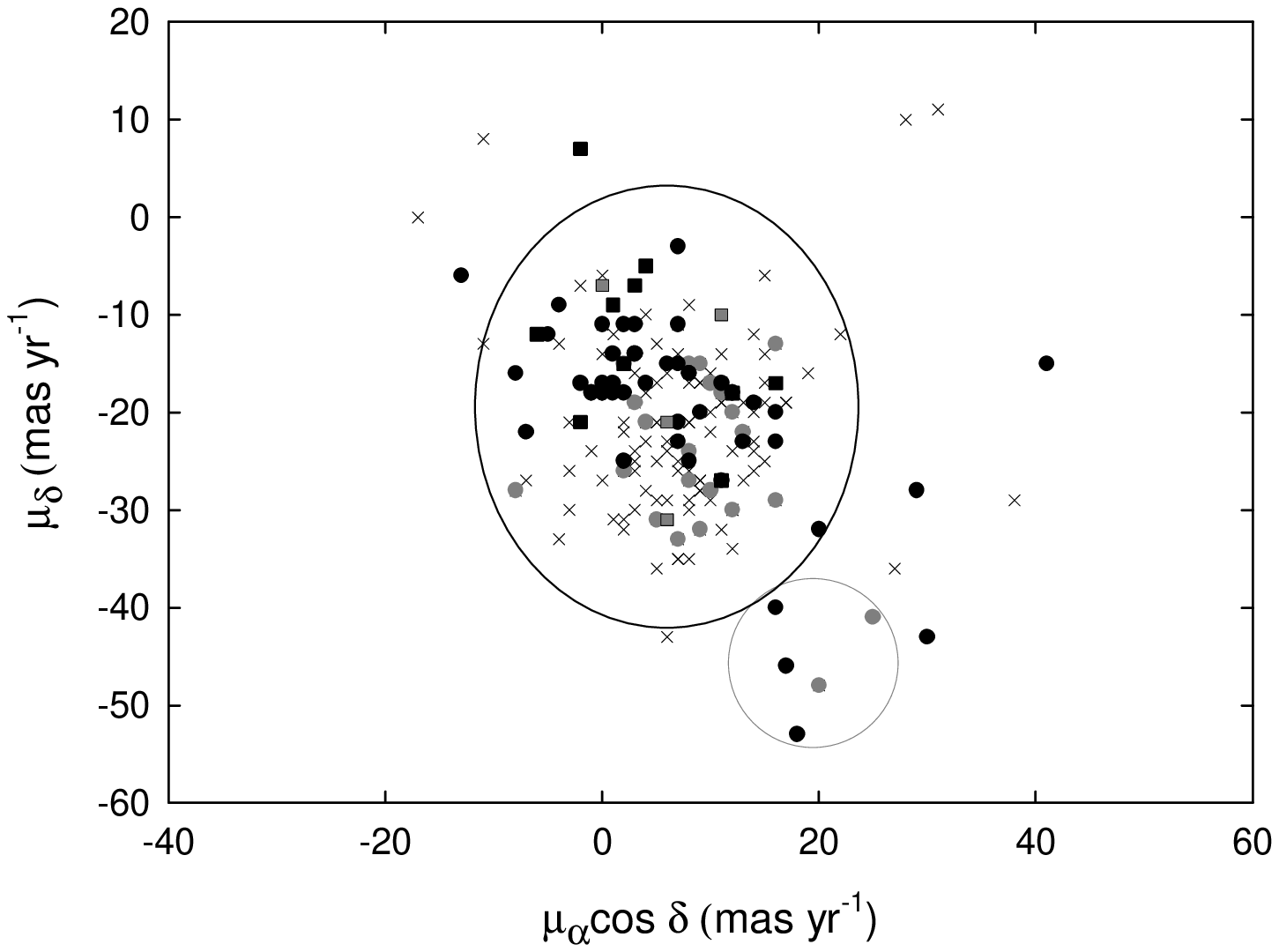}
\caption{\rm \footnotesize {Proper motions for 23 well-known PMS and 48 
PMS candidates from our sample (the gray and black circles, 
respectively). The recognized members of the Taurus--Auriga SFR from 
the list by Kenyon et al. (2008) are represented by the crosses. The 
large ellipse bounds the 3$\sigma$ region around the mean value of the 
proper motions for 123 recognized SFR members. Five stars with proper 
motions close to those of the Pleiades members are enclosed in the 
small gray circle. Fourteen stars without reliable data on their 
luminosities, radii, masses, and ages are designated by the gray 
(4 PMS) and black (10 PMS candidates) squares.}}
\end{figure}

\subsection*{LITHIUM LINE EQUIVALENT WIDTH}
\indent

In Fig. 2, the lithium equivalent width (EW(Li))
is plotted against the effective temperature ($T_{eff}$)
for the objects from our sample and the Pleiades
stars. EW(Li) for the Pleiades stars were taken
from Soderblom et al. (1993). The gray and black
circles designate, respectively, 25 well-known PMS
and 45 PMS candidates from our sample for which
we were able to find EW(Li). The Pleiades stars
are represented by the crosses. The dashed curve is
the upper envelope for the sample of stars from the
Pleiades. It is used as the boundary between the
PMS and ZAMS stars (with an age of $\sim$100 Myr).
If an object from our sample is located above the
enveloping dashed line, then we deem it to be a fairly
young object, with an age of less than 100 Myr.

It can be seen from the figure that 16 stars from
our sample lie in the region of the diagram where the
Pleiades stars are located. At the same time, a significant 
fraction of our objects (54 of 70) are in the region
of young PMS stars. In addition to the main group of
70 stars, we marked 12 more objects without reliable
data on their luminosities, radii, masses, and ages
(see Table 2) for which EW(Li) are known. These are
designated by the gray (4 PMS) and black (8 PMS
candidates) squares. Seven stars from this group are
located in the region of relatively old Pleiades stars
(W9, W19, W49, W50,W61, W72, and TAP~49).

\begin{figure}[ht]
\epsfxsize=10cm
\vspace{0.6cm}
\hspace{3cm}\epsffile{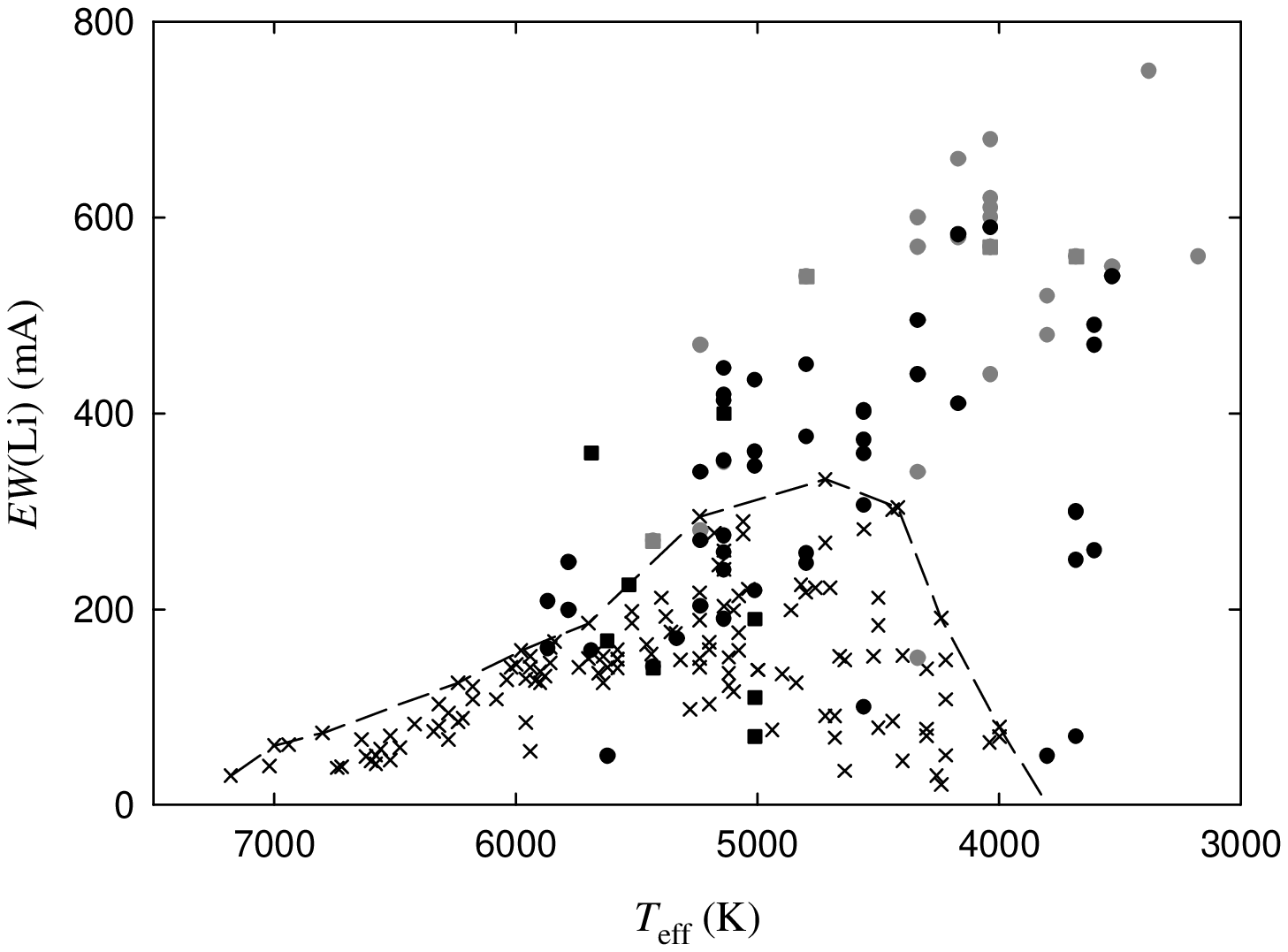}
\caption{\rm \footnotesize {EW(Li) versus $T_{eff}$ for the Pleiades 
(crosses), 25 well-known PMS (gray circles), and 45 PMS candidates 
(black circles). Twelve objects without reliable data on their 
luminosities, radii, masses, and ages are designated by the gray 
(4 PMS) and black (8 PMS candidates) squares. The dashed curve 
indicates the upper envelope for the sample of stars from the
Pleiades.}}
\end{figure}

\begin{figure}[ht]
\epsfxsize=6cm
\vspace{0.6cm}
\hspace{5cm}\epsffile{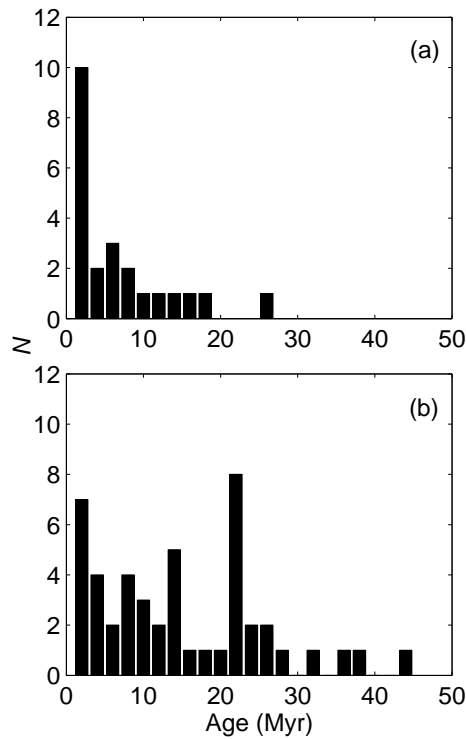}
\caption{\rm \footnotesize {Age distribution for the well-known 
PMS stars (a) and PMS candidates from Wichmann’s list (b).}}
\end{figure}

\subsection*{AGE}
\indent

Previously (Grankin 2013a), we estimated
the ages of the sample stars from the Hertzsprung-
Russell (HR) diagram using evolutionary tracks from
Siess et al. (2000). Figure 3 shows the age distribution
(histogram) for the well-known PMS stars
from our sample (Fig. 3a) and the PMS candidates
(Fig. 3b). We excluded four objects with ages older
than 70 Myr from consideration. The evolutionary
status of these four objects is discussed in the next
section.

It can be seen from the figure that the ages of the
well-known PMS stars are, on average, younger than
those of the PMS candidates. The age distribution
for the well-known PMS is spiky, with its peak near
2 Myr. About half of the well-known PMS objects
have an age of about 2 Myr; the remaining stars have
ages in the range 4-18 Myr. The distribution for the
PMS candidates is wider with three low peaks near 2,
14, and 22 Myr. In the next section, we refine the evolutionary
status of the sample stars using their ages, lithium and 
$H\alpha$ line equivalent widths, and proper motions.

\subsection*{EVOLUTIONARY STATUS}
\indent

As has been shown in the preceding section, there
are young objects with ages of 1-10 Myr both in the
group of well-known PMS stars and in the group
of PMS candidates from Wichmann’s list. There is
good reason to suppose that such young objects are
members of the Taurus--Auriga SFR. However, quite
a few objects with ages of 10-20 Myr are present in
both groups and there are even objects with ages of
20-45 Myr in the group of candidates for PMS stars.
To classify the objects under study and to refine their
evolutionary status, we invoked data not only on their
ages but also on their proper motions and the lithium
and $H\alpha$ line equivalent widths.

As a result of our analysis of the available data,
we propose the following classification scheme. We
classify all objects from our sample with ages up to
10 Myr as WTTS candidates and those with ages of
11-50 Myr as PTTS candidates. An object has a
very reliable evolutionary status, WTTS or PTTS, if:
(1) it exhibits a fairly strong lithium line (lies above the
region occupied by the Pleiades members in Fig. 2),
(2) its $H\alpha$ line is in emission, and (3) its proper motion
corresponds to that of the Taurus--Auriga SFR.
We designate such an object as “wtts+” or “ptts+”,
depending on its age.

If one of the parameters does not meet the requirements
described above or if there are no data for it,
then we deem such an object to have a fairly reliable
evolutionary status, WTTS or PTTS, and designate
it as “wtts” or “ptts”, depending on its age. If any two
parameters do not meet the requirements described
above or if there are no data for them, then we deem
such an object to have an unreliable evolutionary
status and designate it as “wtts?” or “ptts?”. If
none of the parameters, except the age, does not meet
the requirements described above or if there are no
data for all of these parameters, except the age, then
we deem such an object to have a very unreliable
evolutionary status and designate it as “wtts??” or
“ptts??”.

Before reaching conclusions about the evolutionary
status of the candidates for PMS stars from our
sample, we tested the proposed classification scheme
on the group of well-known PMS stars from our
sample. All 24 well-known PMS stars enter into the
catalog by Herbig and Bell (1988) and were classified
as young weak-lined objects (WTTS). According to
the proposed classification scheme, 14 well-known
PMS objects are very reliable WTTS (“wtts+”), two
objects are marked as reliable WTTS (“wtts”), and
two objects are possible WTTS (“wtts?”). Two objects
were classified as very reliable PTTS (“ptts+”),
two objects were marked as reliable PTTS (“ptts”),
one object is a possible PTTS (“ptts?”), and one object
(TAP~4) belongs to the Pleiades (for a discussion,
see below).

Thus, 16 well-known PMS stars from our sample
are reliable WTTS (“wtts+” and “wtts”) and 4 stars
are reliable PTTS (“ptts+” and “ptts”). If we turn to
the most complete list of young stars in the Taurus--
Auriga SFR published by Kenyon et al. (2008), then
it can be noted that all 16 objects that we assign
to reliable WTTS are present in this list, while the
remaining four objects are absent. In other words,
the classification scheme proposed here yields reliable
results for the objects from our sample with a known
evolutionary status, and we can apply this classification
to the group of candidates for PMS stars from our
sample whose evolutionary status has been actively
discussed in recent papers (see the Introduction).

According to the proposed classification scheme,
28\% (14 of 50) of the PMS candidates are reliable
WTTS (“wtts+” and “wtts”) and 32\% (16 of 50) of
the PMS candidates are reliable PTTS (“ptts+” and
“ptts”). The remaining 34\% (17 of 50) have an unreliable
evolutionary status. Thus, it can be asserted that
60\% of the PMS candidates from Wichmann’s list
are reliable PMS stars (WTTS and PTTS) with ages
of 1-40 Myr that can be members of the 
Taurus--Auriga SFR with a high probability. In general, our
conclusions are consistent with those of Wichmann
et al. (2000) with some exceptions. For example,
among the 29 common stars designated by Wichmann
et al. (2000) as PMS, we classified 28 objects
as reliable WTTS and PTTS. Only one object (W15)
has a contradictory classification. In contrast to
Wichmann et al. (2000), we believe that it belongs to
the ZAMS. Similar conclusions can also be reached
about the objects that were classified as stars with
a controversial evolutionary status. For example,
among the 15 common stars designated by Wichmann
et al. (2000) as ZAMS objects, we classified
12 stars as unreliable WTTS and PTTS. Only three
objects (W07, W29, and W54) have a contradictory
classification. In contrast to Wichmann et al. (2000),
we classified these three stars as reliable WTTS. Our
classification of these three stars is consistent with
that of Martin and Magazz\`u (1999) and Bertout and
Genova (2006).

The results of the proposed classification of the
objects under study are presented in Table~3. Several
objects do not fit into the proposed scheme because
of their old ages. We are talking about W15, W39,
W46, and TAP~4. The age of the first three objects is
close to 100 Myr. Without any doubt, they belong to
zero-age main-sequence (ZAMS) stars. We designated
their status as “ZAMS”. TAP~4 have a proper
motion and age like those of the Pleiades stars. In
addition, this star is close in space to the Pleiades
and is characterized by a very rapid rotation (P =
0.482 day). We designated its evolutionary status as
“Pleiades”.

To summarize, it can be said that there are 50
reliable PMS stars (30 WTTS and 20 PTTS) among
the stars of our sample. The axial rotation periods
are known for 35 objects from this group. At the
same time, there are 20 stars with an unreliable evolutionary
status, with the exception of three ZAMS
objects and one object from the Pleiades. Wichmann
et al. (2000) supposed that the objects with
an unreliable evolutionary status could represent the
population of older stars from the Gould Belt.

Here, we point out, however, that many properties
of the 20 stars that have no reliable evolutionary status
within our classification scheme are very similar to
those of the 50 PMS stars with a reliable evolutionary
status. All 20 stars with an unreliable evolutionary
status are located toward the Taurus--Auriga SFR;
their proper motions and age estimates correspond
to those of the recognized SFR members. All these
stars were discovered owing to their enhanced X-ray
luminosity (see Tables 1 and 2) suggesting a solar-type
activity. Indeed, 15 of the 20 stars with an unreliable
evolutionary status exhibit the phenomenon of
spotted rotational modulation, just as 35 stars from
the group of reliable members of the Taurus--Auriga
SFR. Using the known rotation periods and photometric
data, we estimated the mean distance to these
15 stars (for the technique, see Grankin 2013a). It is
$110\pm33$ pc and agrees satisfactorily with the mean
distance to the Taurus--Auriga SFR ($\simeq 140$ pc). All
these circumstances allow the two subgroups of stars
(with a reliable and unreliable evolutionary status) to
be considered as one combined group of 70 young,
magnetically active pre-main-sequence stars located
toward the Taurus--Auriga SFR. In the next section,
we analyze the possible relationship between the rotation
period, mass, and age of these PMS objects.

\begin{table}[t]
\vspace{6mm}
\caption{Evolutionary status of the sample objects}
\centering
\label{meansp} 
\vspace{6mm}
\begin{footnotesize}

\begin{tabular}{c|l||r|l}
\hline\hline
\rule{0pt}{1pt}&&&\\
    W96/name &   Status &  W96/name &     Status \\[2pt]
\hline
\rule{0pt}{1pt}&&&\\
        01 &      wtts? &        60 &      ptts?? \\
        02 &      ptts+ &        62 &      ptts+ \\
        03 &      ptts? &        63 &      wtts+ \\
        04 &      ptts  &        64 &      ptts  \\
        05 &      ptts  &        66 &      ptts? \\
        06 &      ptts  &        67 &      ptts? \\
        07 &      wtts+ &        68 &      ptts? \\
        08 &      ptts? &        70 &      wtts+ \\
        10 &      ptts  &        71 &      ptts \\
        11 &      wtts+ &        73 &      wtts \\
        12 &      wtts  &        74 &      ptts+ \\
        13 &      ptts+ &        75 &      wtts \\
        14 &      ptts? &        76 &      ptts+ \\
        15 &      ZAMS  &     Anon 1 &      wtts \\
        18 &      ptts? &     TAP 31 &      wtts? \\
        23 &      ptts+ &     LkCa 1 &      wtts+ \\
        27 &      ptts  &     LkCa 4 &      wtts+ \\
        29 &      wtts  &     LkCa 5 &      wtts+ \\
        30 &      ptts? &     LkCa 7 &      wtts+ \\
        31 &      ptts??&    LkCa 14 &      wtts+ \\
        32 &      ptts? &    LkCa 16 &      wtts+ \\
        36 &      ptts? &    LkCa 19 &      ptts+ \\
        37 &      wtts+ &    LkCa 21 &      wtts+ \\
        39 &      ZAMS  &      TAP 4 &   Pleiades \\
        40 &      wtts+ &      TAP 9 &      ptts  \\
        41 &       ptts &     TAP 26 &      ptts+ \\
        44 &       ptts &     TAP 35 &      ptts? \\
        45 &      wtts+ &     TAP 40 &      ptts  \\
        46 &       ZAMS &     TAP 41 &      wtts+ \\
        47 &      ptts+ &     TAP 45 &      wtts+ \\
        48 &      wtts  &     TAP 50 &      wtts? \\
        53 &      wtts??&     TAP 57 &      wtts+ \\
        54 &      wtts  &   V819 Tau &      wtts+ \\
        56 &      wtts? &   V826 Tau &      wtts  \\
        57 &      ptts? &   V827 Tau &      wtts+ \\
        58 &      ptts? &   V830 Tau &      wtts+ \\
        59 &      wtts+ &     VY Tau &      wtts+ \\
\hline
\end{tabular}  
\end{footnotesize}
\end{table}

\clearpage
\subsection*{ROTATION PERIOD, MASS, AND AGE}
\indent

Since the magnetically active stars in Taurus--Auriga have quite 
different masses, it is interesting
to investigate the possible rotation period-mass relationship.
We took the rotation periods, masses,
and ages from our previous paper (Grankin 2013a).
The sample of stars with known rotation periods was
divided into four subgroups, depending on the mass.
The condition that the number of objects be the same
in all subgroups was met. Figure 4 presents the
distributions (histograms) of rotation periods for each
individual subgroup of stars. It can be seen from the
figure that the stars with the lowest masses ($0.24<M<0.8 M_\odot$) 
exhibit rotation periods in the entire
range of detected periods. The stars from the second
subgroup ($0.8<M<1.0M_\odot$) have predominantly
short rotation periods, no more than 4 days. Only
three stars from this subgroup exhibit periods in the
range from 4 to 10 days. The third subgroup ($1.0<M<1.2M_\odot$) 
contains only one star with a period
longer than 4 days. Finally, the relatively massive
stars ($1.2<M<1.9M_\odot$) exhibit only short rotation
periods, no more than 3 days. Thus, stars with short
rotation periods are present in all four subgroups irrespective
of the mass. In contrast, the number of stars
with long rotation periods ($P_{rot} > 4$) decreases
appreciably with increasing mass of the objects. For
this reason, the more massive PMS stars from our
sample rotate, on average, faster than the less massive
ones. The mean rotation periods for the four
subgroups as the mass increases are the following:
4.13, 3.06, 2.16, and 1.97 days, respectively.

\begin{figure}[ht]
\epsfxsize=7.0cm
\vspace{0.6cm}
\hspace{4.5cm}\epsffile{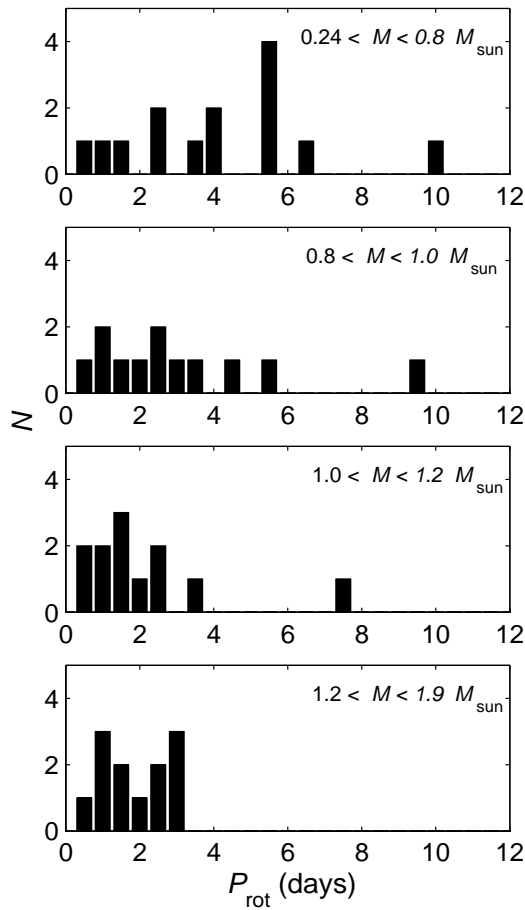}
\caption{\rm \footnotesize {Distributions of rotation periods 
(histograms) for four subgroups of PMS stars with different masses.}}
\end{figure}

Figure 5 presents the distributions (histograms)
of rotation periods for two different age subgroups of
magnetically active stars: for 25 WTTS ($t< 10$ Myr)
and 25 PTTS ($t > 10$ Myr). It can be seen from
the figure that the younger PMS stars (WTTS) have
both short (\mbox{$P<5$} days) and long rotation periods in
the range from 5 to 10 days. In contrast, the older
PMS stars (PTTS) with ages of more than 10 Myr
exhibit only short rotation periods. In other words,
the relatively old PMS stars with ages of more than
10 Myr rotate faster than the youngest stars with ages
of less than 10 Myr. This result is in good agreement
with the theoretical models that predict an increase
in the rotation velocity (a decrease in the rotation
period) as the PMS stars move toward the ZAMS. In
the next section, we discuss in detail the evolution of
angular momentum for the magnetically active stars
from our sample.

\begin{figure}[h]
\epsfxsize=7.0cm
\vspace{0.6cm}
\hspace{5.0cm}\epsffile{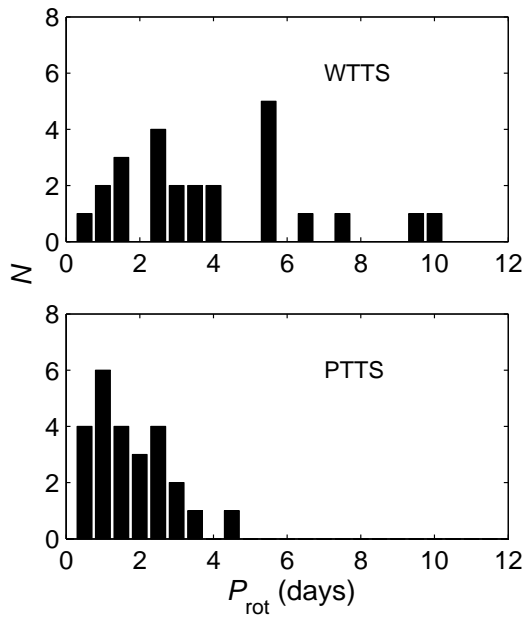}
\caption{\rm \footnotesize {Distributions of rotation periods for 25 WTTS
($t< 10$ Myr) and 25 PTTS ($t > 10$ Myr).}}
\end{figure}

\vspace{-2pt}
\subsection*{THE EVOLUTION OF ANGULAR MOMENTUM}
\indent

The spin-up of solar-type stars on PMS radiative
tracks is believed to result from a decrease in the
moment of inertia. As soon as the PMS stage ends,
the moment of inertia subsequently remains essentially
constant and a magnetized wind brakes the star
during its MS evolution. Within the first 5-10 Myr
of evolution, the star-protoplanetary disk interaction
is believed to be the main cause of the angular momentum
loss by PMS stars. The exact mechanism
responsible for the angular momentum loss is still
being actively discussed. At present, the two most
probable mechanisms are considered: the removal
of angular momentum through a stellar wind (see,
e.g., Matt and Pudritz 2005) and an active interaction
between the stellar magnetic field and the ionized gas
in the inner protoplanetary disk regions (see, e.g.,
Konigl 1991; Collier Cameron and Campbel 1993).

It is more preferable to use the rotation periods
than the equatorial rotation velocities to effectively
test particular models for the evolution of angular
momentum, because the rotation periods are measured
directly and are more accurate than the rotation
velocities, which are affected by the projection effect.

Apart from the rotation periods of the stars from
our sample, we used the rotation periods for solar-type
dwarfs (with masses of 0.9-1.1$M_\odot$) in the
clusters IC 2602 (an age of 30 Myr), Alpha Persei
(50 Myr), the Pleiades (100 Myr), and the Hyades
(700 Myr) from Pizzolato et al. (2003). The rotation
period is plotted against the age for the stars from our
sample and for dwarfs from the clusters listed above in
Fig. 6. To interpret the observational data, we applied
a simple model that we had already used previously
(see Bouvier et al. 1997) and that is described in detail
in Bouvier (1994) and Bouvier and Forestini (1994).
This model describes the presumed evolution of the
rotation period for a solar-mass star as it evolves from
the T Tauri phase to the Sun’s age. We made the
following assumptions when calculating this model:

(1) The initial rotation period is 8 days, as is observed
in many young T Tauri stars.

(2) Solid-body rotation of the star ($d\Omega/dr=0$)
during its PMS and MS evolution.

(3) The existence of a magnetospheric interaction
between the star and the disk: all the time the star is
surrounded by the disk, it evolves with a constant angular
velocity ($d\Omega/dt=0$, i.e., $V(t)=V_oR(t)/R_o$).

(4) The evolution of the star’s angular momentum
under the action of a stellar wind after the loss of its
disk: the braking law $dJ/dt\propto\ \Omega^2$ for rapidly rotating
stars ($> 8$ km $\rm s^{-1}$) and $dJ/dt\propto\ \Omega^3$ for slowly rotating
stars ($\leq 8$ km $\rm s^{-1}$).

\begin{figure}[h]
\epsfxsize=11.0cm
\vspace{0.6cm}
\hspace{3cm}\epsffile{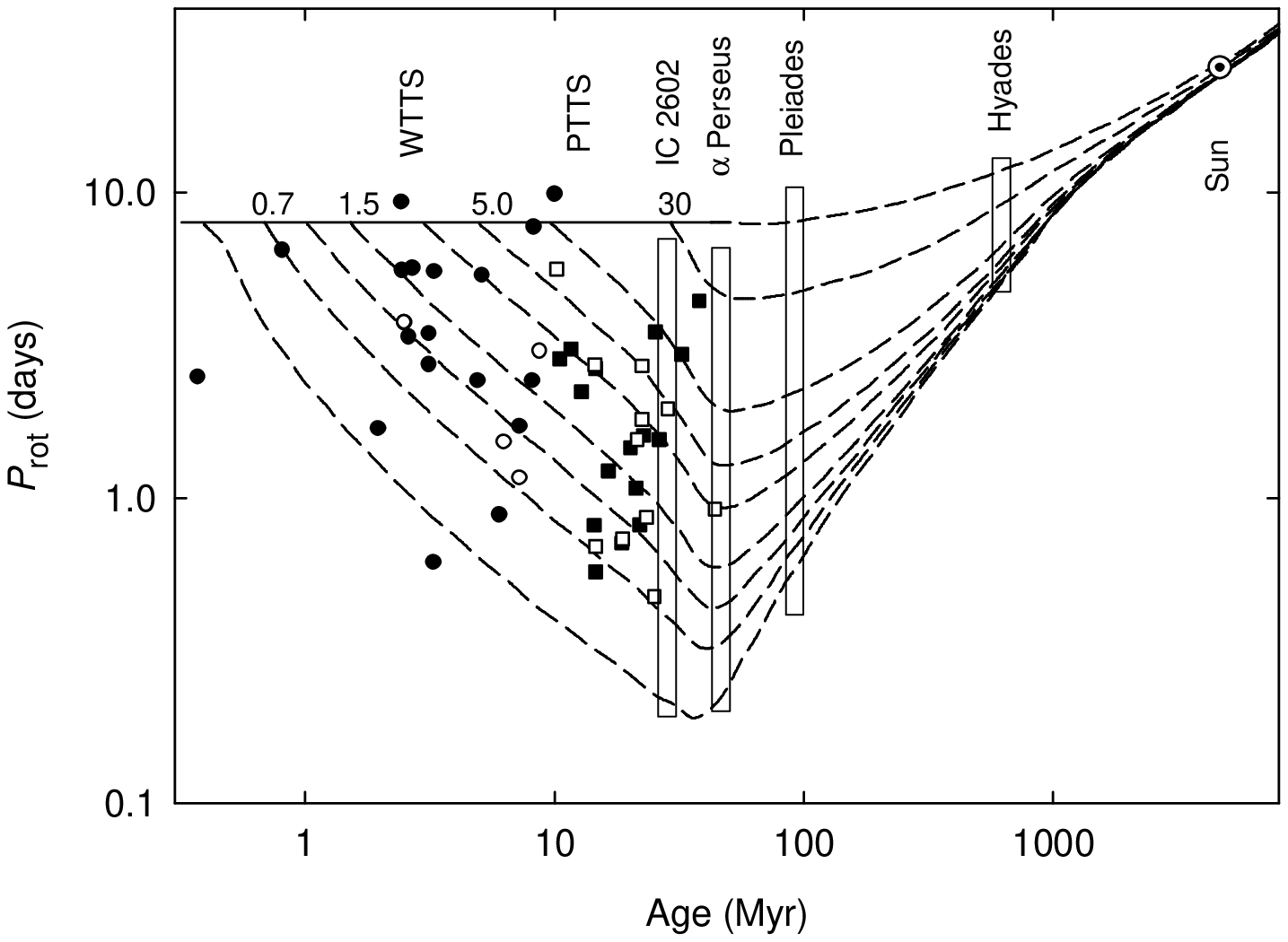}
\caption{\rm \footnotesize {Rotation period versus age. WTTS with a 
reliable and unreliable evolutionary status are designated by the black and
white circles, respectively. PTTS with a reliable and unreliable 
evolutionary status are marked by the black and white squares,
respectively. The light rectangles indicate the positions of cluster 
dwarfs with various ages. The upper solid horizontal line was
calculated for a star rotating with a constant period of 8 days and 
interacting with the protoplanetary disk during the entire
PMS stage (50 Myr). Each dashed line indicates the evolution of the 
stellar rotation period depending on the specific instant
of time when the magnetospheric connection with the disk is lost. These 
correspond to the following instants of time: 0.4, 0.7,
1.0, 1.5, 3.0, 5.0, 10, 30, and 50 Myr.}}
\end{figure}

At present, it is commonly assumed that young
stars evolve on their convective tracks without spin-up
during all the time they remain in interaction with
the protoplanetary disks. Each dashed curve shown
in Fig. 6 starts at a certain time of the star’s PMS
evolution: the starting point of the curve corresponds
to the age when the magnetospheric connection between
the star and its disk ceases. Thereafter, the star
begins to rotate freely, remaining under the action of
only the braking wind, which carries away part of the
angular momentum.

The first dashed curve in Fig. 6 starts at an age of
0.4 Myr. At this time, the solar-mass star is slightly
below the so-called birth line in the region of the HP
diagram where CTTS are actually observed (Kenyon
and Hartmann 1995). The starting points of the
remaining dashed lines shown in Fig. 6 correspond
to ages of 0.7, 1.0, 1.5, 3.0, 5.0, 10, 30, and 50 Myr,
respectively.

There is a large scatter of rotation periods for all
ages for both WTTS and PTTS. According to the
proposed model, this scatter can be explained by
different active star-disk interaction times. Most of
the stars from our sample lie above the track corresponding
to an age of 0.7 Myr and below the track
corresponding to an age of 10 Myr. It is within this
time interval that an active star-protoplanetary disk
interaction occurs.

Thus, the observed spin-up from the youngest
WTTS to PTTS within the time scale from 1 to
40 Myr is well explained by a rapid decrease in the
moment of inertia, while the angular momentum
losses are moderate. It should be noted that there
also exist other models based on completely different
ideas that also predict spin-up for stars on PMS
radiative tracks (Soderblom et al. 1993; Keppens
et al. 1995; Cameron et al. 1995). These models
are based on the idea of the so-called saturated
dynamo suggesting the absence of an active core-shell 
interaction instead of uniform internal rotation.

\subsection*{CONCLUSIONS}
\indent

We analyzed a sample of 74 magnetically active
stars toward the Taurus--Auriga SFR. Based on accurate
data on their basic physical parameters obtained
from original photometric observations and
published data on their proper motions, X-ray luminosities,
and equivalent widths of the $H\alpha$ and Li lines,
we refined the evolutionary status and analyzed the
relationship between the rotation, mass, and age of
these objects. In particular, the following results were
obtained.

We showed that 16 well-known PMS objects from
our sample are reliable WTTS and 4 PMS objects
are reliable PTTS that belong to the Taurus--Auriga
SFR. Analysis of the evolutionary status for 50 candidates
for PMS stars from Wichmann’s list showed
that 60\% (30 of 50) of the objects are young stars
with ages of 1-40 Myr (14 WTTS and 16 PTTS)
that could belong to the Taurus--Auriga SFR with a
high probability. We classified 34\% (17 of 50) of the
candidates for PMS stars from Wichmann’s list as
stars with an unreliable evolutionary status that could
be assigned to the population of Gould Belt stars.
The remaining three stars with ages of about 100 Myr
were classified as ZAMS objects.

As a result of this work, we identified a group of
70 young (1-40 Myr), magnetically active stars with
reliable data on their rotation periods, luminosities,
radii, and masses toward the Taurus--Auriga SFR.
The dependence of the rotation period on mass and
age was investigated for 50 PMS stars. We showed
that the more massive stars of the sample rotate,
on average, faster than the less massive ones. The
relatively old PMS stars (PTTS, t \textgreater 10 Myr) rotate,
on average, faster than the younger ones (WTTS,
t \textless 10 Myr). This result is in good agreement with
the theoretical models that predict a decrease in the
rotation period when moving toward the ZAMS.

We investigated the evolution of angular momentum
for PMS stars in Taurus--Auriga. We showed
that the distribution of stars on the rotation period-age 
diagram could be explained by different active
star-disk interaction times. Most of the sample stars
ceases to actively interact with their disks on a time
scale from 0.7 to 10 Myr. The observed spin-up
from the youngest WTTS to PTTS within a time
scale from 1 to 40 Myr is well explained by a rapid
decrease in the moment of inertia, while the angular
momentum losses are moderate.

The possible relationship between various magnetic
activity parameters and rotation for 70 PMS
stars toward the Taurus-Auriga SFR will be investigated
in the next paper.


\subsection*{REFERENCES}

\begin{enumerate}

\item G. Basri, E. L. Martin, and C. Bertout, Astron. Astrophys.
\textbf{252}, 625 (1991).

\vspace{-1ex}
\item C. Bertout and F. Genova, Astron. Astrophys. \textbf{460},
499 (2006).

\vspace{-1ex}
\item J. Bouvier, ASP Conf. Ser. \textbf{64}, 151 (1994).

\vspace{-1ex}
\item J. Bouvier and M. Forestini, in \textit{Circumstellar Dust
Disk and Planetary Formation, Proceedings of the 10th IAP Meeting}, 
Ed. by P. Ferlet (Frontieres, Gif-sur-Yvette, 1994), p. 347.

\vspace{-1ex}
\item J. Bouvier, R. Wichmann, K. Grankin, et al., Astron.
Astrophys. \textbf{318}, 495 (1997).

\vspace{-1ex}
\item A. C. Cameron, C. G. Campbell, and H. Quaintrell,
Astron. Astrophys. \textbf{298}, 133 (1995).

\vspace{-1ex}
\item A. C. Cameron and C. G. Campbel, Astron. Astrophys.
\textbf{274}, 309 (1993).

\vspace{-1ex}
\item J.M. Carpenter, Astron. J. \textbf{120}, 3139 (2000).

\vspace{-1ex}
\item Duy Cuong Nguyen, Ray Jayawardhana,
Marten H. van Kerkwijk, et al., Astrophys. J. \textbf{695},
1648 (2009).

\vspace{-1ex}
\item C. Ducourant, R. Teixeira, J. P. P\'eri\'e, et al., Astron.
Astrophys. \textbf{438}, 769 (2005).

\vspace{-1ex}
\item E. D. Feigelson, Astrophys. J. \textbf{468}, 306 (1996).

\vspace{-1ex}
\item S. Frink, S. R\"ooser, R. Neuh\"auser, et al. Astron.
Astrophys. \textbf{325}, 613 (1997).

\vspace{-1ex}
\item M. Gomez, B. F. Jones, L. Hartmann, et al., Astron.
J. \textbf{104}, 762 (1992).

\vspace{-1ex}
\item K. N. Grankin, S. Yu. Melnikov, J. Bouvier, et al.,
Astron. Astrophys. \textbf{461}, 183 (2007).

\vspace{-1ex}
\item K. N. Grankin, J. Bouvier, W. Herbst, et al., Astron.
Astrophys. \textbf{479}, 827 (2008).

\vspace{-1ex}
\item K. N. Grankin, Astronomy Letters \textbf{39}, (4), 251 (2013a).
\vspace{-1ex}
\item M. G\"udel, K. R. Briggs, K. Arzner, et al. Astron.
Astrophys. \textbf{468}, 353 (2007).

\vspace{-1ex}
\item L.W. Hartmann, D. R. Soderblom, and J. R. Stauffer,
Astron. J. \textbf{93}, 907 (1987).

\vspace{-1ex}
\item G. H. Herbig, in \textit{Can Post-T Tauri Stars Be Found?},
Ed. by L. V. Mirzoyan (Publ. House Armenian Acad.
Sci., Yerevan, 1978), p. 171.

\vspace{-1ex}
\item G. H. Herbig, F. J. Vrba, and A. E. Rydgren, Astron.
J. \textbf{91}, 575 (1986).

\vspace{-1ex}
\item G. H. Herbig and K. R. Bell, \textit{Third Catalog of
Emission Line Stars of the Orion Population}, Lick
Obser. Bull. (Lick Observ., Santa Cruz, 1988).

\vspace{-1ex}
\item S. J. Kenyon and L. Hartmann, Astrophys. J. Suppl.
Ser. \textbf{101}, 117 (1995).

\vspace{-1ex}
\item S. J. Kenyon, M. G\'omez, and B. A. Whitney, in
\textit{Handbook of Star Forming Regions}, Vol. 1: \emph{The
Northern Sky}, Ed. by Bo Reipurth (ASP Monograph
Publ., 2008), vol. 4, p. 405.

\vspace{-1ex}
\item R. Keppens, K. B. MacGregor, and P. Charbonneau,
Astron. Astrophys. \textbf{294}, 469 (1995).

\vspace{-1ex}
\item A. Konigl, Astrophys. J. \textbf{370}, L39 (1991).

\vspace{-1ex}
\item A. Magazz\`u, R. Rebolo, and Y. V. Pavlenko, Astrophys.
J. \textbf{392}, 159 (1992).

\vspace{-1ex}
\item E. L. Martin, R. Rebolo, A. Magazz\`u, et al., Astron.
Astrophys. \textbf{282}, 503 (1994).

\vspace{-1ex}
\item E. L. Martin, Astron. Astrophys. \textbf{321}, 492 (1997).

\vspace{-1ex}
\item E. L. Martin and A. Magazz\`u, Astron. Astrophys.
\textbf{342}, 173 (1999).

\vspace{-1ex}
\item S. Matt and R. E. Pudritz, Astrophys. J. \textbf{632}, L135
(2005).

\vspace{-1ex}
\item R. Neuh\"auser, G. Torres, M. F. Sterzik, et al., Astron.
Astrophys. \textbf{325}, 647 (1997).

\vspace{-1ex}
\item F. Palla and S. W. Stahler, Astrophys. J. \textbf{540}, 255
(2000).

\vspace{-1ex}
\item N. Pizzolato, A. Maggio, G. Micela, et al., Astron.
Astrophys. \textbf{397}, 147 (2003).

\vspace{-1ex}
\item L. Siess, E. Dufour, and M. Forestini, Astron. Astrophys.
\textbf{358}, 593 (2000).

\vspace{-1ex}
\item D. R. Soderblom, B. F. Jones, S. Balachandran, et al.,
Astron. J. \textbf{106}, 1059 (1993).

\vspace{-1ex}
\item B. Stelzer and R. Neuh\"auser, Astron. Astrophys. \textbf{377},
538 (2001).

\vspace{-1ex}
\item K. M. Strom, F. P. Wilkin, S. E. Strom, et al., Astron.
J. \textbf{98}, 1444 (1989).

\vspace{-1ex}
\item F. M.Walter, A. Brown, R. D. Mathieu, et al., Astron.
J. \textbf{96}, 297 (1988).

\vspace{-1ex}
\item R. Wichmann, J. Krautter, J.H.M.M. Schmitt, et al.,
Astron. Astrophys. \textbf{312}, 439 (1996).

\vspace{-1ex}
\item R. Wichmann, G. Torres, C. H. F. Melo, et al., Astron.
Astrophys. \textbf{359}, 181 (2000).
\end{enumerate}

\end{document}